\documentclass[a4paper]{jpconf}
\usepackage{graphicx}

\usepackage{bm,bbm,euscript}
\usepackage{amstext,amsmath,amssymb,amsfonts}
\usepackage{epsfig}
\usepackage{verbatim}
\usepackage{latexsym}

\def\1{{{\mathbbm 1}}}
\def\6{\langle}
\def\9{\rangle}
\newcommand{\ket}[1]{\left | #1 \right\rangle}

\def\tr{{\rm tr}\,}

\def\half{\mbox{$1\over2$}}
\def\quar{\mbox{$1\over4$}}

\def\be{\begin{equation}}
\def\ee{\end{equation}}

\def\bay{\begin{array}}
\def\ear{\end{array}}
\def\arr{\rightarrow}

\def\tr{{\mathrm{tr}}\,}

\def\eL{\EuScript{L}}

\newcommand{\Ecal}{\mathcal{E}}
\newcommand{\Hcal}{\mathcal{H}}
\newcommand{\Ical}{\mathcal{I}}
\newcommand{\Pcal}{\mathcal{P}}

\def\mrin{{\mathrm{in}}}

\begin{document}

\title{Quantum discord in quantum computation}

\author{Aharon Brodutch$^{1}$,
Alexei Gilchrist$^{1}$,
Daniel R. Terno$^{1}$, and Christopher J. Wood$^{2}$}
\address{
$^1$Centre for Quantum Computer Technology, Department of Physics and Astronomy, Macquarie University, Sydney NSW 2113, Australia

$^2$ Institute for Quantum Computing, University of Waterloo, 200 University Av. West
Waterloo Ontario N2L 3G1, Canada}

\ead{aharon.brodutch@mq.edu.au}

\begin{abstract}
Quantum discord is a  measure of the quantumness of correlations. After reviewing its different versions and properties, we apply it to the questions of quantum information processing. First we show that changes in discord in the processed unentangled states indicate the need for entanglement in the distributed implementation of quantum gates. On the other hand, it was shown that zero system-environment discord is a necessary and sufficient condition for applicability of the standard completely positive description of the system's evolution. We demonstrate that this result  does not translate into useful quantum process tomography.  Depending on the details of the preparation procedure  only absence of any initial correlations may guarantees consistency of the process tomography.
\end{abstract}

\section{Introduction}
What makes correlations quantum? What makes quantum computer tick? There is no definite answer to these questions, even if entanglement is a primary suspect. It is crucial in performing tasks that cannot be implemented by local operations and classical communications (LOCC),   its presence in the system is a natural telltale that something non-local is aground, and one entangling gate is a part of universal sets of gates \cite{NielsenandChuang,BrusLeuchs}.

Nevertheless, the story is not so simple. The ``sausage" or ``four and a half tatami teahouse" states \cite{Bennettetal99}  are nine orthogonal pure product states that are shared between the two parties --- A(lice) and B(ob). Despite  absence of entanglement Alice and Bob cannot distinguish between them  by LOCC.  Similarly the DQC1 algorithm demonstrates speed up without entanglement \cite{DattaShajiCaves08}.

Quantum discord \cite{OllivierZurek01,HendersonVerdal01} is one of the  measures of quantum correlations. We review its three versions and their properties in the next Section.  Since its introduction it was found to be related to the efficiency of various Maxwell's demons \cite{Zurek03,BTdiscord}, quantum phase transitions \cite{Dillenschneider08,Mazieroetal10b}, and quantum state merging \cite{MadhokDatta10,Cavalcantietal10}. It was also argued that zero discord is a necessary and sufficient condition for the evolution of a system to be of a completely positive type \cite{Rodrigiuezetal08,lidar}, thus being amenable to a relatively simple characterization,  in quantum tomography.

We present  two of its uses. First,
 ability to create entanglement is necessary for execution of two-qubit quantum gates even when they are applied to unentangled  states and create no entanglement. Changes in discord indicate the failure of a local implementation. Second,  having a zero system-environment discord does not guarantee that the resulting tomographic description is an adequate representation of the relevant gate action. Hence zero discord is not directly related to a  ``simple" evolution of the system.

\section{Three types of discord} \label{Secdiscord}

\subsection{Correlations and entropies}
The Shannon entropy \cite{tc} of a classical discrete  probability distribution  over a random variable $A$ with values $a$, $p(a)\equiv p_a$, is defined by $H(A)=-\sum_a p_a\log p_a$.  Entropy of the joint probability distribution  over $AB$, $H(AB)$,  is defined analogously.  The Bayes theorem relates joint and conditional probabilities,
\be
p(a,b)=p(a|b)p(b)=p(b|a)p(a),
\ee
where $p(a|b)$ is a conditional probability of $A=a$ given that $B=b$. The conditional entropy of $A$ given $B$,
\be
H(A|B)=\sum_b p_b H(A|b)=-\sum_{a,b}p(a,b)\log p(a|b)
\ee
is a weighted average of the entropies of $A$ given a particular outcome of $B$.

Correlations between  two probability distributions are measured by the symmetric mutual information. It has two equivalent expressions \cite{tc},
\be
I(A:B)=H(A)+H(B)-H(A,B) \label{mi} \; ;
J(A:B)=H(A)-H(A|B)=H(B)-H(B|A). 
\ee

Quantum-mechanical (von Neumann) entropy \cite{NielsenandChuang,BrusLeuchs,peres} of a system with a density operator $\rho$  is defined as $S(\rho)=-\tr\rho\log\rho$. It minimizes the Shannon entropy of  probability distributions that result from rank-1 positive operator-valued measures (POVMs) that are applied to the state $\rho$. The minimum is  reached on a probability distribution $A_\rho^\Pi$ that results from a projective measurement $\Pi=\{\Pi_a, a=1,\ldots d\}$, $\sum_a\Pi_a=\1$, $\Pi_a\Pi_b=\delta_{ab}\Pi_a$, which is constructed from the eigenstates of $\rho$,
\be
S(\rho)=\min_\Pi H(A_\rho^\Pi) \label{minent},
\ee
i. e.,
\be
S(\rho)=H(A_\rho^{{\Pi^*}}), \qquad \rho=\sum_a p_a{\Pi}_a^*, \qquad p_a\geq 0,\quad \sum_a p_a=1.
\ee

The first expression for mutual information has an obvious quantum generalization,
\be
I(\rho_{AB}):=S(\rho_A)+S(\rho_B)-S(\rho_{AB}),
\ee
and represents the total amount of quantum and classical correlations \cite{Horodeckietal05}.

To obtain a quantum version of $J(A:B)$, it is necessary to determine a conditional state of the subsystem $B$ \cite{BTdiscord}.
Given a complete projective measurement $\Pi$ on $A$, a quantum definition of $J$ follows its interpretation as the information gained about the system $B$ from the measurement on $A$ \cite{ OllivierZurek01},
\be \label{Jdef}
J^{\Pi^A}(\rho_{AB}):=S(\rho_B)-S(\rho_B|\Pi^A),
\ee
where the conditional entropy is now given by $
S(\rho_B|\Pi^A):=\sum_a p_a S(\rho_{B|\Pi_a})$. 

The post-measurement state of $B$ that  corresponds to the outcome $A=a$  is
\be
\rho_{B|\Pi_a}= (\Pi_a\otimes\1_B \rho_{AB}\Pi_a\otimes\1_B)/p_a, \qquad p_a=\tr\rho_A\Pi_a. \label{def-b},
\ee
while the state of $B$ remains unchanged, $\rho_B=\tr_A\rho_{AB}=\sum_a p_a\rho_{B|\Pi_a}$.


\subsection{Discords $D_1$ and $D_2$}
Unlike their classical counterparts, the quantum expressions are generally inequivalent and $I(\rho_{AB})\geq J^{\Pi^A}(\rho_{AB})$ \cite{ OllivierZurek01,HendersonVerdal01}. The  difference between these two quantities is
\be
D_1^{\Pi^A}(\rho_{AB}):=S(\rho_A)+S(\rho_B|\Pi^A)-S(\rho_{AB}).
\ee
Its dependence on the measurement procedure is removed by minimizing the result over all possible sets of $\Pi$, resulting in the quantum discord \cite{OllivierZurek01}
\be\label{D1def}
D_1^A(\rho_{AB}):=\min_{\Pi^A} D_1^{\Pi^A}(\rho_{AB}).
\ee
Similarly,
\be
J_1^A(\rho_{AB}):=\max_{\Pi^A}J^{\Pi^A}(\rho_{AB}).
\ee
 It is possible to define the discord when the difference is minimized over all possible  POVM $\Lambda^A$ \cite{HendersonVerdal01}. However, unless stated otherwise we restrict ourselves to rank 1 projective measurements. 

An explicit form of a post-measurement state will be useful in the following text. We denote this state as $\rho'_{X}\equiv\rho^{\Pi^A}_X$, where the subscript $X$ stands for $A$, $B$, or $AB$, and  use the former expression if it does not lead to  confusion. After a projective measurement $\Pi^A$, the state of the system becomes
\be
\rho_{AB}'=\sum_a p_a\Pi_a\otimes\rho^a_B,  \label{postmes}
\ee
where $p_a$ and $\rho_B^a\equiv\rho_{B|\Pi_a}$ are given by Eq.~\eqref{def-b}.

The discord of the state $\rho_{AB}$ is zero if and only if it is a mixture of products of arbitrary states of $B$ and projectors on $A$ \cite{OllivierZurek01},
\be
\rho_{AB}=\sum_a p_a\Pi_a\otimes\rho^a_B, \qquad  p_a\geq 0, \quad \sum_a p_a=1. \label{zerodisc}
\ee

By using this decomposition and properties of the entropy of block-diagonal matrices \cite{wehrl} we can identify $
J^{\Pi^A}(\rho_{AB})\equiv I(\rho_{AB}'),
\label{mutinfo}$
because $S(\rho_A')=H(A_\rho^\Pi)$ and
\be
S(\rho_{AB}')=H(A_\rho^\Pi)+S(\rho_B|\Pi^A). \label{ent-post}
\ee
The discord is not a symmetric quantity: it is possible to have states  with $D_1^A(\rho_{AB})\neq D_1^B(\rho_{AB})$.

Another possibility  is to set
\be
{J}_2^{\Pi^A}:=S(\rho_A)+S(\rho_B)- [H(A_\rho^\Pi)+S(\rho_B|\Pi^A)]=S(\rho_A)+S(\rho_B)-S(\rho_{AB}'),
\ee
arriving to the quantum discord as defined in \cite{Zurek03}
\be
D_2^A(\rho_{AB}):=\min_\Pi[H(A_\rho^\Pi)+S(\rho_B|\Pi^A)]-S(\rho_{AB}), \label{def_d2}
\ee
where the quantity to be optimized is a sum of post-measurement entropies of $A$ and $B$.
Using Eq.~\eqref{minent} we see that $D_1\leq D_2$. It is also easy to see that $D_1=0 \Leftrightarrow D_2=0$. Using Eqs.~\eqref{postmes} and \eqref{ent-post} we obtain a different expression for $D_2$:
\be
D_2^{\Pi^A}(\rho_{AB})=S(\rho_{AB}^{\Pi^A})-S(\rho_{AB}). \label{ent_inc2}
\ee

Since the definition of the discord(s) involves  optimization, the analytic expressions are known only in some particular cases. Moreover, typically it is important to know whether the discord is zero or not, while the numerical value itself is less significant.

\subsection{Zero discord and $D_3$}
It follows from Eq. \eqref{zerodisc}  that if the spectrum of a reduced state $\rho_A=\sum_a p_a\Pi_a$ is non-degenerate, then its  eigenbasis gives a unique family of projectors $\Pi$ that results in the zero discord for $\rho_{AB}$. Hence, a recipe for testing states for zero discord and for finding the optimal basis is to trace out a subsystem that is left alone ($B$),  to diagonalize $\rho_A$ and to calculate the discord in the resulting eigenbasis.

If the state $\rho_A$ is degenerate, a full diagonalization should be used. For the state with the form of Eq.~\eqref{zerodisc},  each of the reduced states $\rho^a_B$ can be diagonalized as
\be
\rho^a_B=\sum_b r^a_b P^b_a, \qquad P^b_a P^{b'}_a=\delta^{bb'}P^b_a.
\ee
The eigendecomposition of the state $\rho_{AB}$ then easily follows. Writing it as
\be
\rho_{AB}=\sum_{a,b}w_a r^a_b \Pi_a\otimes P^b_a, \label{two0}
\ee
it is immediate to see that its eigenprojectors are given by $\Pi_a\otimes P^b_a$. Hence, if $\rho_B$  has a degenerate spectrum, but $\rho_{AB}$ does not, the structure of its eigenvectors reveals if it is of a zero or nonzero discord. Hence we established \cite{BTdiscord}

\textbf{Property 1.}
The eigenvectors of a zero discord state $D_1^A(\rho_{AB})=0$ satisfy
\be
\rho_{AB}|ab\9=r_{ab}|ab\9, \quad\Rightarrow\quad |ab\9\6ab|=\Pi_a\otimes P^b_a.
\ee
$~$ \hfill $\Box$

\noindent This consideration leads to the simplest necessary condition for zero discord \cite{Ferraroetal10}:

\textbf{Property 2.} If $D_1^A(\rho_{AB})=0$, then
$[\rho_A\otimes\1_B,\rho_{AB}]=0.$
Hence a non-zero commutator implies $D_1^A(\rho_{AB})>0$. \hfill $\Box$

Naturally, if the state has  zero discord, and the eigenbasis is only partially degenerated, we can use it to reduce the optimization space. On the other hand, the diagonalizing basis  $\Pi_*$ is not necessarily the optimal basis $\hat{\Pi}$ or  $\check{\Pi}$ that enters the definition of $D_1$ or $D_2$, respectively.  Consider, for example, a two-qubit state
\be
\rho_{AB}=\quar(\1_{AB}+b \sigma^z_{A}\otimes\1_B+c\sigma^x_{A}\otimes\sigma^x_{B}),\label{ex_disc}
\ee
where $\sigma^a_X$ are Pauli matrices on the relevant spaces, $X=A,B$, and the constants $b$ and $c$ are restricted only by the requirements that $\rho_{AB}$ is a valid density matrix. For this state $\rho_B=\1/2$ and $\rho_A=\mathrm{diag}(1+b,1-b)/2$. After the measurement in the  diagonalizing basis $\Pi^z=((\1+\sigma^z)/2,(1-\sigma^z)/2)$ the conditional state of $B$ becomes
\be
\rho_{B|\Pi^z_\pm}=\1/2,
\ee
and the conditional entropy is maximal, $S(\rho_B|\Pi^z)=\log 2$.

On the other hand, in the basis $\Pi^x=((\1+\sigma^x)/2,(1-\sigma^x)/2)$ the probabilities of the outcomes are equal, $p_+=p_-=1/2$, but the post-measurement states of $B$ are different from the maximally mixed one,
\be
\rho_{B|\Pi^x_\pm}=\half(\1\pm c \sigma^x),
\ee
so the entropy $S(\rho_B|\Pi^z)\geq S(\rho_B|\Pi^x)$.

This discrepancy motivates us to  define  a new version of the discord \cite{BTdiscord}:
\be
D_3^{A}(\rho_{AB}):=S(\rho_A)-S(\rho_{AB})+S(\rho_B|\Pi^A_*),
\ee
where $\Pi^A_*$ is the set of eigenprojectors of $\rho_{A}$. For the degenerate case it can be introduced using the continuity of entropy in finite-dimensional systems \cite{wehrl}.
By applying Eq.~\eqref{minent} to the subsystem $A$ we find that  $D_3$ simultaneously holds the analog of Eq.~\eqref{mutinfo},
\be
J_3(\rho_{AB})\equiv I(\rho_{AB}^{\Pi_*^A}),
\ee
and of Eq.~\eqref{ent_inc2},
\be
D_3(\rho_{AB})\equiv S(\rho_{AB}^{\Pi^A_*})-S(\rho_{AB}).
\ee

We also arrive to the following ordering of the discord measures:
\be
D_1^A\leq D_2^A\leq D_3^A \label{duneq}.
\ee
There are several important cases when the  measures of discord coincide. Most importantly, they vanish simultaneously:

\textbf{Property 3.} $D_1=0\Leftrightarrow D_2=0\Leftrightarrow D_3=0$.

\noindent The proof follows from Eqs.\eqref{zerodisc} and \eqref{duneq}. \hfill $\Box$.

For pure states the discord is equal to the degree of entanglement,
\be
D_i^A(\phi_{AB})=S(\phi_A)=E(\phi_{AB}), \qquad i=1,2,3.
\ee
Discord is also independent of the basis of measurement if the state is invariant under local rotations \cite{OllivierZurek01}. Finally,
if $A$ is in a maximally mixed state, then $D_1^A=D_2^A$.

These coincidences make it interesting to check when the discords $D_1$ and $D_2$ are different. By returning to the measurement-dependent versions of the discords, we see that
\be
D_1^{\Pi^A}(\rho_{AB})=D_2^{\Pi^A}(\rho_{AB})-[H(A^{\Pi}_\rho)-S(\rho_A)].
\ee
Assume that  $D_2^{\Pi^A}(\rho_{AB})$ reaches the minimum on the set of projectors $\check{\Pi}$, which are not the eigenprojectors of $\rho_A$. In this case $H(A^{\check{\Pi}}_\rho)-S(\rho_A)>0$, so we can conclude that the strict inequality $D_1^A<D_2^A$ holds, because
\begin{align}
D_1^A(\rho_{AB})&\leq D_1^{\check{\Pi}}(\rho_{AB}) \\
=&D_2^A(\rho_{AB})-[H(A^{\check{\Pi}}_\rho)-S(\rho_A)]<D_2^A(\rho_{AB})\nonumber.
\end{align}
For example, the state of Eq.~\eqref{ex_disc}, with $b=c=\half$, satisfies $D_1^A\approx 0.05$, $D_2^A\approx 0.20$ and $D_3^A\approx 0.21$.

\subsection{Some applications}
The discord $D_1^A(\rho)$ has an operational meaning through quantum state merging \cite{MadhokDatta10,Cavalcantietal10}. Different restrictions on the local Maxwell demons that operate on the subsystems, as compared to the power of a global demon, lead to the differences in maximal work extraction that are determined by the discords $D_2^A$ and $D_3^A$ \cite{Zurek03,BTdiscord}.  Further work on the discord in quantum open systems can be found in \cite{decoherence}.

Since zero discord is thought to represent absence of quantum correlations, it is interesting to investigate the following question. Consider a set of pure orthogonal bipartite  states, each of which may have a different prior probability, with the ensemble density matrix $\rho_{AB}$.  Does the  value of $D(\rho_{AB})$ tell us something about the ability to perfectly distinguish these states by local operations and classical communication (LOCC)? The answer is negative \cite{BTdiscord}: there is no relation  between $D(\rho_{AB})$ and  local distinguishability, as illustrated in Table 1. We use the nine teahouse states
\be
|1\9\otimes |1\9, \quad |0\9\otimes|0\pm 1\9/\sqrt{2},\quad |2\9\otimes|1\pm2\9/\sqrt{2},
|1\pm2\9\otimes|0\9/\sqrt{2}, \quad |0\pm 1\9|2\9/\sqrt{2},
\ee
of \cite{Bennettetal99}, and the result \cite{wv00} that any two orthogonal (entangled or not) states can be perfectly distinguished by LOCC to compile the table.

\begin{table}[htbp]
   \centering
    \caption{Local measurability vs. discord}     \label{Discordtab}
    \begin{tabular}{llc}
   \br

     ~~~~~~~~~~~~~~~~States    & Discord & Locally Distiguishable\\
      \mr
     9 teahouse states, equal weights   & $D^A=D^B=0$ & no  \\
          2 product bi-orthogonal states   & $D^A=D^B=0$     &  yes \\
         2  entangled orthogonal  states  & $D_1^A>0$  & yes  \\
     9 teahouse states, unequal weights    & $D_1^A>0$  & no \\
     \hline
   \end{tabular}
   \label{tab:booktabs}
\end{table}

\section{Restricted quantum gates}
 It is usually thought that quantum computers are potentially faster then their classical counterparts because of the ability to create and use entanglement during the computation \cite{JozsaLinden2003}. Nevertheless, recent results indicate that it is discord rather than entanglement which is responsible for the speed-up \cite{DattaShajiCaves08,BTgates,Eastin10,Fanchinietal10}.

Using a simple example we show how changes in discord indicate the need for entanglement as a resource even if the processed states are unentangled. More general results may be found in \cite{BTgates}. The non-classicality of quantum gates can be examined by the amount of quantum resources required for the operation of the gate. Our method is to analyze the resources that are needed for a bi-local implementation of a gate.

A CNOT gate  can be implemented  bi-locally by Alice and Bob if they share one ebit of entanglement per gate use \cite{Eisert2000}. If they  can perform only LOCC they cannot implement this gate even on a restricted set of unentagled inputs that are transformed into unentagled outputs. One such set  $\eL$  is in Table \ref{input4}.

\begin{table}[htbp]
   \centering
    \caption{Four inputs/outputs for the CNOT gate}     \label{input4}
\begin{tabular}[v]{cl|cl}\hline\hline
\# & State & \# & {State} \\ \hline
$a$&$|1\9|Y_+\9 \arr i|1\9|Y_-\9$ & $c$ &$|Y_+\9|X_-\9 \arr |Y_-\9|X_-\9$\\
$b$&$|0\9|Y_+\9 \arr|0\9|Y_+\9$ & $d$ & $|Y_+\9|X_+\9 \arr |Y_+\9|X_+\9$\\\hline\hline
\end{tabular}
\end{table}

\noindent Here $\sigma_y|Y_\pm\9=\pm|Y_\pm\9$, $\sigma_x|X_\pm\9=\pm|X_\pm\9$, where $\sigma_{x,y,z}$ are Pauli matrices.

We show that the ability to implement the CNOT gate on $\eL$ without shared entanglement allows one to  unambiguously discriminate between these non-orthogonal states using just one input copy, which is impossible \cite{peres}. Without specifying the local operations of Alice and Bob we classify them  according to their action on the state $|Y_+\9$. An operation  $\Phi$ is \textit{flipping} (F) if up to a phase $\Phi(|Y_+\9)=|Y_-\9$, \textit{non-flipping} (N) if  $\Phi(|Y_+\9)=|Y_+\9$, and undetermined otherwise.

Knowing the type of the operation allows  Alice and Bob to narrow down the list of the possible inputs: e.g., Bob's F is incompatible with having input $b$, while for Alice's operation not to have a definite type excludes both $c$ and $d$.  If one of the operations done is neither F nor N, then the type of other operations allows to determine the input uniquely.

Any pair of outputs can be reset to the original inputs by local unitaries and resent through the gate. For example, if the overall operation was of FF type, the operation $\sigma_z^A\otimes\sigma_x^B$ transforms the outputs $\psi_a'=|1\9|Y_-\9$ and $\psi_c'=|Y_-\9|X_-\9$ into the inputs $\psi_a$ and $\psi_c$.

The operations that implement the gate this time may be of the same type as before, or different. If the gate is such that there is a finite probability of having a different operation type, it will be realized after a finite number of trials. This other type (FN or NF in the above example) will uniquely specify  the inputs. If the gate's  design is such that  a particular pair of inputs is always processed by the same type of operations, then the gate can be used to unambiguously distinguish between one state from this pair and at least one of the two remaining states in a single trial.  \hfill$\square$

The operation in the  example changes the discord of the ensemble and has been extended in \cite{BTgates} to a more general setting involving a symmetrised type of discord defined as
\be
D_2(\rho):=\min[D_2^A(\rho),D_2^B(\rho)]\neq 0.
\ee
It is however not yet clear if discord is indeed the best measure for this kind of non locality. It is noteworthy to mention  that the algorithm described in  \cite{DattaShajiCaves08} changes the discord of only one of the parties so that $D_2$ remains zero throughout the computation.

\section{(Non) completely positive maps and discord}
\subsection{Completely positive maps and gate tomography}
Environmental interactions are the major obstacle for practical quantum information processing.  Together with the imperfect tailoring of gate Hamiltonians they are responsible for discrepancies between the ideal unitary gates and their experimental realizations. This is why characterization
of quantum processes is an essential step in implementing quantum technology. The exposition below is based on \cite{chris}. Experimental realization of this discussion is currently in progress.

Usually the transformation of input to output states of an open system is described by completely positive (CP) maps. Any such map $\Ecal(\rho)$ can be seen as a reduction of a unitary evolution of some initially uncorrelated system--environment state $\rho_{AB}=\rho_A\otimes\omega_B$, where the initial states of the system and the environment were $\rho_A$ and $\omega_B$, respectively.  Under the joint evolution $U_{AB}$ the state of the system is transformed to
\be
\rho'_A\equiv\Ecal(\rho_A)=\tr_B U\rho_{AB} U^\dag=\sum_a M_a\rho_A M_a^\dag, \label{cpdef}
\ee
where $M_a$ are the Kraus representation matrices \cite{NielsenandChuang,BrusLeuchs}. Using the spectral
decomposition $\omega=\sum_\nu p_\nu|\nu\rangle\langle\nu|$ one recovers the Kraus matrices from

\begin{equation}
\rho'=
 \sum_{\mu,\nu}\langle\mu|\sqrt{p_\nu}\, U|\nu\rangle\rho
 \langle\nu|\sqrt{p_\nu}\,U^\dagger|\mu\rangle.
\label{envir}
\end{equation}
This result is based on the absence of prior correlations between the system and the environment. 

The action of a given gate is reconstructed using several methods of quantum process tomography. We focus on the standard process tomography, which consists in following. A set of input  states $\{\rho_j\}$ is prepared and sent through the process $\mathcal{E}$. From the knowledge of the input  and the reconstructed outputs $\Ecal(\rho_k)$ states the process matrix $\chi$  \cite{NielsenandChuang, dhk, chris09} is reconstructed. If $K_m$, $m=1,\ldots, d^2$ form the basis for operators acting on $\Hcal_A$, $d=\dim \Hcal_A$, then
\be
\Ecal(\rho)=\sum_{m,n}\chi_{mn}K_m\rho K_n^\dag.
\ee
For a CP evolution all the eigenvalues of $\chi$ are non-negative, and the Kraus matrices are obtained as its (generalized) eigenvectors.

Various mathematical techniques are used to convert the relative frequencies of different measurement outcomes  into the states $\Ecal(\rho_k)$. Positivity of states $\rho'_j$ and complete positivity of the process $\Ecal$ are enforced, and its violations by the raw data are interpreted as a result of  noise.

The combined state of a system  ($A$) and its environment ($B$)
can be represented  in the {Fano form}
\be
 \rho_{AB}=\frac{1}{d_Ad_B}\Bigl( \1_{AB}+\sum_i\alpha_i\sigma^A_i
 \otimes\1_B
 +\sum_j\beta_j\1_A\otimes\sigma^B_j
 +\sum_{ij}\gamma_{ij}\sigma^A_i\otimes\sigma^B_j \Bigr).\label{Fano}
\ee
Here the $\sigma_i^X$, $i=1,...,d_X^2$ represent generators of SU($d_X$),
while the real vector $\vec \alpha$ (or $\vec \alpha$) of size $d_X^2-1$
is the generalized Bloch vector of the reduced density operator
$\rho_X$.
The correlations between subsystems $A$ and $B$
are characterized by
\begin{equation}
 \Gamma_{ij}=(\gamma_{ij}-\alpha_i\beta_j)/d_Ad_B.\label{corrtens}
\end{equation}
Presence of correlations not only can lead to a non-CP evolution of the system, but blurs its
boundary with the environment. Part of the controversy surrounding non-CP maps in literature
\cite{dhk} can be traced to this ambiguity as well as to the impossibility of unrestricted creation of arbitrary input states of the system (that are to form a tomographically complete basis in the standard CP paradigm) $\rho_{j}$ \textit{and} their correlations with the environment $\Gamma_{j}$.

This situation has an important bearing on the process tomography and action of quantum processing devices. In the following we show that while zero discord $D_A(\rho_{AB})$ ensures a complete positivity of the resulting evolution \cite{Rodrigiuezetal08}, it does not necessarily translate into a useful tomographic description of the ensuing evolution.

\subsection{Role of preparations}
State preparation is a vital aspect
of quantum process tomography. When
initial correlations are present in the input state, the preparation procedure used has considerable impact on the outcome.  We
discuss two such schemes: state preparation by a measurement and rotations and state preparation by projective
measurements \cite{Preparation}.

In the former case (Method 1) a single post-selected state that corresponds to a projection  operator $\rho_H=\Pi_H$ is used.
Once we have performed the projective measurement the required input states are obtained by applying the appropriate matrix from a set of unitary rotation matrices $\{ R_{a}\}$ such that $R_{a} \rho_H R_{a}^{\dagger}=\rho_{a}$ where $\rho_{a}$ is run over the tomographically complete set of input states.

Hence the preparation maps for Method 1 are given by
\be
\Pcal_a(\rho_{AB}^\mrin)
= \frac{1}{p_H}\left(R_a\Pi_H^A\otimes\1_B\right) (\rho_{AB}^\mrin) \left(\Pi_H^A R^\dagger_a\otimes\1_B\right)= \rho_a\otimes\omega_H,  \label{eqn:preprotations}
\ee
where $p_H=\tr(\Pi_H^A\rho_A^\mrin)$ is the probability of detecting the outcome corresponding $H$, and $R_j$  brings $|H\9\6H|$ to one of the  states $\rho_a\equiv|\psi_a\9\6\psi_a|$. The post-preparation state of the environment is dependent on both the initial state $\rho_{AB}^\mrin$ \emph{and} the measurement operator $\Pi_H$.

State preparation using \emph{only} measurements (Method 2) utilizes the set of projectors $\{\Pi_a\}$ where $\Pi_a\equiv\rho_a$. Hence the preparation procedure for projective measurements is given by the collection of maps $\{ \Pcal_{a} \}$ where
\be
\Pcal_{a}(\rho_{AB}^\mrin)
= \frac{1}{p_a}\left( \Pi_{a}^A \otimes\1_B\right)\rho_{AB}^\mrin \left(\Pi_{a}^A\otimes\1_B\right)    = \rho_{a}\otimes\omega_{a}\label{eqn:prepmeas}.
\ee
where $p_a=\tr(\Pi_a^A\rho_A^\mrin)$ is the probability of detecting the outcome corresponding to $\Pi_j$.
In this preparation scheme  the state of the environment, $\omega_j$ depends on the probe state of the system, $\rho_k$, hence  the process matrix will not necessarily be CP.

\textbf{Lemma 1:} Zero discord of the initial state does not guarantee a CP evolution in a tomographic procedure.

Consider the state
\be
 \rho_{AB}^\mrin=\frac{1}{2}\left(|0\9\6 0|\otimes|0\9\6 0| + |1\9\6 1|\otimes|1\9\6 1| \right),
\ee
which satisfies even a stronger requirement $D^A(\rho_{AB}^\mrin)=D^B(\rho_{AB}^\mrin)=0$. If the tomographic procedure is based on preparations by measurements, the resulting process matrix $\chi$ has the  eigenvalues $(1\pm\sqrt{3}/2, \pm\sqrt{3}/2)$ and hence the evolution is not CP. \hfill $\square$

Preparation by measurement and rotations is guarantied to lead to a CP evolution. However, we should be concerned that the actual state on which the gate will operate is different from the ones prepared in this procedure, and hence the predictive power of the resulting process matrix $\chi$ will be low. Having a zero discord does not alleviate this difficulty, as shown below. We consider different reconstruction of process matrices for a generic fixed overall unitary  $U_{AB}$.

\textbf{Lemma 2:} Reconstructed process matrices $\chi_{\Pcal_a}$ that follow from the spanning set of measurement preparations $\{\Pcal_a^A\}$ that are given by projectors $\Pi_a^A$, $a=1,\ldots d^2_A-1$, applied to the same initial system--environment state $(\rho_{AB}^\mrin)$, are equal  if and only if this initial state  is simply separable, $(\rho_{AB}^\mrin)=(\rho_{A}^\mrin)\otimes(\omega_{B}^\mrin)$.

The sufficient part is obvious. To prove the necessary condition we recall that Kraus matrices form an eigendecomposition of the process matrix, hence  from Eq.~\eqref{envir} it follows that all post-selected environmental states $\omega_k$ have to have the same spectrum. Hence up to a local reshuffling  after the preparation by measurement $\Pi_a$ 
\be
\rho_{AB}^\mrin\mapsto \Pcal_{a}^A(\rho_{AB}^\mrin)=\Pi_a\otimes\omega_B^0.
\ee
Using Fano decomposition of Eq.~\eqref{Fano} we find that for all $a$ the post-measurement state of $B$ is
\be
\omega_B^0=\tr_A\Pcal_{a}^A(\rho_{AB}^\mrin)=\frac{1}{d_B}\big(\1_B+\sum_j\beta_j^\mrin\sigma_j^B\big)+\frac{1}{p_a}
\sum_j\left(\sum_i\Gamma_{ij}^\mrin\tr(\Pi_a\sigma_i^A)\right)\sigma_j^B,
\ee
hence for any $j$
\be
\sum_i\Gamma_{ij}^\mrin\6\psi_a|\sigma_i|\psi_a\9=0,
\ee
for all $a$ in all sets of tomographically complete projectors. As a result, $\Gamma_{ij}^\mrin=0$ so the initial state $\rho_{AB}^\mrin$ is indeed a direct product $\rho_A^\mrin\otimes\omega_B^0$. \hfill $\square$

\subsection{Examples}
We illustrate the differences between the outcomes of Methods 1 and 2 by considering qubits as the system and environment and a CNOT gate with system A as the target being the
overall unitary. Let  the initial state of joint system $AB$ to be  the maximally entangled state $\rho_{\Phi^+}=|{\Phi^+}\9\6\Phi^+|$ where $\ket{\Phi^+}=(\ket{00}+\ket{11})/\sqrt{2}$.

The reconstructed process matrices are
\be
\chi_1=
        \left(
            \begin{array}{cccc}
             1 & 0 & 0 & 1 \\
             0 & 0 & 0 & 0 \\
             0 & 0 & 0 & 0 \\
             1 & 0 & 0 & 1
            \end{array}
        \right) , \quad
        \chi_2=\frac{1}{2}
        \left(\begin{array}{ c c c c }
 			2 & 0 &-1-i & 1   \\
			0 & 0 & 1 & 1+i \\
			-1+i & 1 & 2 & 0 \\
    		1 & 1-i & 0 & 0 \end{array}
        \right),
\ee
for Methods 1 and 2, respectively.
 As follows from Eq.~\eqref{eqn:preprotations}, in the former case we have a CP evolution (actually it is an identity  $\Ecal_1=\Ical$), while
$\chi_2$ on the other hand has eigenvalues $(1+\frac{\sqrt{3}}{2},-\frac{\sqrt{3}}{2},\frac{\sqrt{3}}{2},1-\frac{\sqrt{3}}{2})$ and hence the  reconstructed process  is not completely positive.
To see why this happens  consider the state of the environment after each preparation procedure. We have that all states are different,
\be
\omega_H=			\left(\begin{array}{ c c }
 							1 & 0   \\
    							0 & 0 \end{array} \right), \
\omega_V=   		\left(\begin{array}{ c c }
 							0 & 0   \\
    							0 & 1 \end{array} \right),\
\omega_D=			\frac{1}{2}\left(\begin{array}{ c c }
 							1 & 1   \\
    							1 & 1\end{array} \right),\
\omega_R=	 	\frac{1}{2}	\left(\begin{array}{ c c }
 							1 & i   \\
    							-i & 1 \end{array} \right).
\ee
While we used a maximally entangled input state, this result holds even for separable inputs. For example, for the input state $\rho_{AB}^\mrin=\frac{1}{2}\left(  \rho_H\otimes\rho_A + \rho_D\otimes\rho_V   \right)$,
the  process matrix $\chi_2$ has the eigenvalues $\{1.642, 0.507, -0.253, 0.105\}$, so the evolution is still non-CP.

\vspace{10pt}
We would like to thank C. Rodriguez-Rosario and K. Modi for their helpful comments and suggestions.

\end{document}